\documentclass[aps,prd,preprint,superscriptaddress,11pt]{revtex4-1}

\usepackage{amsmath}
\usepackage{graphics}
\usepackage{graphicx}
\usepackage[mathcal]{eucal}

\begin{document}

\preprint{COLO-HEP-562, UCI-TR-2011-12}

\title{Leptogenesis in a SUSY $SU(5) \times T^{\prime}$ Model with Geometrical CP Violation}

\author{Mu-Chun Chen}
\email[]{muchunc@uci.edu}
\affiliation{Department of Physics \& Astronomy, 
University of California, Irvine, CA 92697-4575, U.S.A.}
\author{K.T. Mahanthappa}
\email[]{ktm@pizero.colorado.edu}
\affiliation{Department of Physics, University of Colorado at Boulder, Boulder, CO 80309-0390, U.S.A.}

%\date{April 9, 2009}

\begin{abstract}
The complex Clebsch-Gordon coefficients in the double tetrahedral group $T^{\prime}$ offers the possibility that CP violation can be entirely geometrical in origin, as pointed out by us recently. In this note, we investigate leptogenesis in a SUSY $SU(5) \times T^{\prime}$ model, which gives rise to realistic quark masses and CKM matrix elements, a near tri-bimaximal neutrino mixing pattern, as well as CP violating measures for all quarks  and leptons. In terms of 9 independent parameters in the Yukawa sector, the predicted values of the 22 observables agree with current experimental values, including the recent T2K and MINOS results. The correction to the tri-bimaximal mixing pattern is related to the Cabibbo angle, leading to interesting quark-lepton complementarity sum rules. Our predicted value for the leptonic Dirac CP phase is $\delta_{\ell}^{\mbox{\tiny CP}} = 227^{o}$, which gives rise to sufficient amount of lepton number asymmetry, in the presence of the flavor effect. As this is the only non-vanishing CP violating phase in the lepton sector, there is a direct connection between leptogenesis and CP violation in neutrino oscillation in our model. 
\end{abstract}

%\keywords{}

\maketitle

\section{Introduction}
\label{sec:intro}

The measurement of various neutrino oscillation parameters has entered a precision era. At present the global fit to a suite of oscillation experiments indicate the following best fit values and $3\sigma$ limits~\cite{Fogli:2011qn}, 
\begin{eqnarray}
\sin^{2} \theta_{atm} =  0.42 \; (0.34 - 0.64) \; , \; 
\sin^{2}  \theta_{\odot}  =  0.306 \; (0.259-0.359) \; , \nonumber \\
\sin^{2} \theta_{13} =  0.021 \; (0.001-0.044) \; , \nonumber \\
\Delta m_{atm}^{2}  =  2.35 \; (2.06-2.67)  \times 10^{-3} \; \mbox{eV}^{2} \; , \; 
\Delta m_{\odot}^{2}  =  7.58 \; (6.99-8.18) \times 10^{-5} \; \mbox{eV}^{2} \; .
\end{eqnarray}   
The experimental values for the neutrino mixing angles are very close to the prediction of the tri-bimaximal mixing (TBM) matrix~\cite{Harrison:2002er},
\begin{equation}
U_{TBM} = \left(\begin{array}{ccc}
\sqrt{2/3} & \sqrt{1/3} & 0 \\
-\sqrt{1/6} & \sqrt{1/3} & -\sqrt{1/2} \\
-\sqrt{1/6} & \sqrt{1/3} & \sqrt{1/2}
\end{array}\right) \; , 
\end{equation}
which predicts
\begin{equation}
\sin^{2} \theta_{atm}^{\mbox{\tiny TBM}} = 1/2 \; , \quad \tan^{2}\theta_{\odot}^{\mbox{\tiny TBM}} = 1/2 \; , \quad \sin\theta_{13}^{\mbox{\tiny TBM}} = 0 \; . 
\end{equation}
The Super Kamiokande (SuperK) Collaboration recently presented~\cite{superK2010} at Neutrino 2010 for the very first time the best fit value for the leptonic Dirac CP phase,
\begin{equation}
\delta_{\ell}^{\mbox{\tiny SK}} = 220^{o} \; .
\end{equation}
The recent result~\cite{Abe:2011sj} from T2K Collaboration has given an indication of non-zero $\theta_{13}$. If the T2K result holds up, it is likely that the value of $\theta_{13}$ will be measured within the next decade by the reactor experiments. In addition, the Long Baseline Neutrino Experiment (LBNE), if approved,  will be able to determine the leptonic Dirac CP violating phase, $\delta_{\ell}$.

It has been realized that the TBM matrix can arise from an underlying $A_{4}$ symmetry~\cite{Ma:2001dn}. Nevertheless, $A_{4}$ does not give rise to quark mixing~\cite{Ma:2006sk}, making it incompatible with grand unified theory (GUT). On the other hand, the group 
$T^{\prime}$~\cite{Chen:2007afa,frampton}, which is the double covering of $A_{4}$, can successfully account for the quark sector as demonstrated in a SU(5) model constructed by us~\cite{Chen:2007afa}. (It is interesting to note that the particle content of Ref.~\cite{Chen:2007afa} is free of discrete gauge anomaly~\cite{Luhn:2008xh,Chen:2006hn}.) One special property of the group $T^{\prime}$ is that its group theoretical Clebsch-Gordon (CG) coefficients are intrinsically complex~\cite{cg}. Based on this observation, we pointed out for the first time in Ref.~\cite{Chen:2009gf} that CP violation can entirely be geometrical in origin due to the complex CG coefficients in $T^{\prime}$.  In this note, we generalize our model to be supersymmetric and investigate the implication of our model for leptogenesis. Details of the vacuum alignment and UV completion of the model will be presented in a separate publication~\cite{CM2011-2}.  

This paper is organized as follows. In Section~\ref{sec:model}, we give the particle content and the Yukawa superpotential of the model. The numerical predictions for the fermion masses and mixing angles as well as the CP violating measures are given in Section~\ref{sec:Numeric}. This is followed by Section~\ref{sec:Leptg} which discusses the implications of leptogenesis in our model. Section~\ref{sec:conclude} concludes the paper.

\section{Model}
\label{sec:model}

The content of the chiral superfields in our model (including the three generations of matter fields, the $SU(5)$ Higgses in the Yukawa sector, and flavon fields) as well as their quantum numbers with respective to $SU(5)$, $T^{\prime}$, and $Z_{12} \times Z_{12}$ are given in Table~\ref{tbl:charge}.  
\begin{table}[t!]
\begin{tabular}{|c|cccc|ccc|cccccc|ccc|}\hline
& $T_{3}$ & $T_{a}$ & $\overline{F}$ & $N$ & $H_{5}$ & $H_{\overline{5}}^{\prime}$ & $\Delta_{45}$ & $\phi$ & $\phi^{\prime}$ & $\psi$ & $\psi^{\prime}$ & $\zeta$ & $\zeta^{\prime} $ & $\xi$ & $\eta$  & $S$ \\ [0.3em] \hline\hline
SU(5) & 10 & 10 & $\overline{5}$ & 1 &  5 & $\overline{5}$ & 45 & 1 & 1 & 1 & 1& 1 & 1 & 1 & 1 & 1\\ \hline
$T^{\prime}$ & 1 & $2$ & 3 & 3 & 1 & 1 & $1^{\prime}$ & 3 & 3 & $2^{\prime}$ & $2$ & $1^{\prime\prime}$ & $1^{\prime}$ & 3 & 1 & 1 \\ [0.2em] \hline
$Z_{12}$ & $\omega^{5}$ & $\omega^{2}$ & $\omega^{5}$ & $\omega^{7}$ & $\omega^{2}$ & $\omega^{2}$ & $\omega^{5}$ & $\omega^{3}$ & $\omega^{2}$ & $\omega^{6}$ & $\omega^{9}$ & $\omega^{9}$ 
& $\omega^{3}$ & $\omega^{10}$ & $\omega^{10}$ & $\omega^{10}$ \\ [0.2em] \hline
$Z_{12}^{\prime}$ & $\omega$ & $\omega^{4}$ & $\omega^{8}$ & $\omega^{5}$ & $\omega^{10}$ & $\omega^{10}$ & $\omega^{3}$ & $\omega^{3}$ & $\omega^{6}$ & $\omega^{7}$ & $\omega^{8}$ & $\omega^{2}$ & $\omega^{11}$ & 1 & $1$ & $\omega^{2}$
\\ \hline   
\end{tabular}
\vspace{-0.in}
\caption{Field content of our model. The three generations of matter fields in $10$ and $\overline{5}$ of $SU(5)$ are in the $T_{3}$, $T_{a}$ $(a=1,2)$ and $\overline{F}$ multiplets. The Higges that are needed to generate $SU(5)$ invariant Yukawa interactions are $H_{5}$, $H_{\overline{5}}^{\prime}$ and $\Delta_{45}$. The flavon fields $\phi$ through $N$ are those that give rise to the charged fermion mass matrices, while $\xi$ and $\eta$ are the ones that generate neutrino masses.  The $Z_{12}$ charges are given in terms of the parameter $\omega = e^{i\pi/6}$.}  
\label{tbl:charge}
\end{table}
This particle content leads to the following Yukawa superpotential up to mass dimension seven,  
\begin{equation}
\mathcal{W}_{\mbox{\tiny Yuk}} =  \mathcal{W}_{TT} + \mathcal{W}_{TF} + \mathcal{W}_{\nu} \; ,
\end{equation}
where  
\begin{eqnarray}
\mathcal{W}_{TT} & = & y_{t} H_{5} T_{3} T_{3} + \frac{1}{\Lambda^{2}}  H_{5} \biggl[ y_{ts} T_{3} T_{a} \psi \zeta + y_{c} T_{a} T_{b} \phi^{2} \biggr] + \frac{1}{\Lambda^{3}} y_{u} H_{5} T_{a} T_{b} \phi^{\prime 3} \quad 
\label{eq:Ltt} \; , \\ 
\mathcal{W}_{TF} & = &  \frac{1}{\Lambda^{2}} y_{b} H_{\overline{5}}^{\prime} \overline{F} T_{3} \phi \zeta + \frac{1}{\Lambda^{3}} \biggl[ y_{s} \Delta_{45} \overline{F} T_{a} \phi \psi \zeta^{\prime}  + y_{d} H_{\overline{5}^{\prime}} \overline{F} T_{a} \phi^{2} \psi^{\prime} \biggr]  \quad   
\label{eq:Ltf} \; , \\
\mathcal{W}_{\nu} & = & \lambda_{1} NNS +  \frac{1}{\Lambda^{3}} \biggl[ H_{5}  \overline{F} N \zeta \zeta^{\prime} \biggl( \lambda_{2} \xi + \lambda_{3} \eta\biggr) \biggr] \quad
\label{eq:Lff} \; .
\end{eqnarray}
The UV completion of these operators is discussed in Ref.~\cite{CM2011-2}. 
Here the parameter $\Lambda$ is the scale above which the $T^{\prime}$ symmetry is exact. The vacuum expectation values of the flavon fields are given by,
\begin{equation}
\left<\xi\right> = \left(\begin{array}{c}
1 \\ 1 \\ 1
\end{array}\right)
\xi_{0} \Lambda \; , \; 
\left< \phi^{\prime} \right> = \left(\begin{array}{c}
1 \\ 1 \\ 1
\end{array}\right) \phi_{0}^{\prime} \Lambda \; , \;  
\left< \phi \right> = \left( \begin{array}{c} 
0 \\ 0 \\ 1
\end{array}\right) \phi_{0} \Lambda \; , \;  
\end{equation}
\begin{equation}
\left< \psi \right> = \left( \begin{array}{c} 1 \\ 0 \end{array}\right)
\psi_{0} \Lambda \; , \; 
\left< \psi^{\prime} \right> = \left(\begin{array}{c} 1 \\ 1 \end{array}\right) \psi_{0}^{\prime} \Lambda \; , 
\end{equation}
\begin{equation}
\left< \zeta \right> = \zeta_{0} \Lambda \; , \; \left< \zeta^{\prime} \right> = \zeta_{0}^{\prime} \Lambda \; , \; 
\left<\eta\right> = \eta_{0} \Lambda \; , \; 
\left<S\right> = s_{0} \Lambda \;  .   
\end{equation}
Note that  all the expectation values are assumed to be real and they don't contribute to CP violation. On the other hand, the reality of the Yukawa coupling constants is ensured as there is sufficient number of flavon fields to absorb the complex phases in the Yukawa coupling constant by field redefinition.

The superpotential gives rise to the following mass matrix for the up-type quarks,
\begin{equation}
M_{u} = \left( \begin{array}{ccc}
i \phi^{\prime 3}_{0}  & (\frac{1-i}{2}) \phi_{0}^{\prime 3} & 0 \\
(\frac{1-i}{2})  \phi_{0}^{\prime 3}  & \phi_{0}^{\prime 3} + (1 - \frac{i}{2}) \phi_{0}^{2} & y^{\prime} \psi_{0} \zeta_{0} \\
0 & y^{\prime} \psi_{0} \zeta_{0} & 1
\end{array} \right) y_{t}v_{u}, \qquad , 
\end{equation}
and the following down-type quark and charged lepton mass matrices,
\begin{eqnarray}
M_{d}  & = & \left( \begin{array}{ccc}
0 & (1+i) \phi_{0} \psi^{\prime}_{0} & 0 \\
-(1-i) \phi_{0} \psi^{\prime}_{0} & \psi_{0} N_{0} & 0 \\
\phi_{0} \psi^{\prime}_{0} & \phi_{0} \psi^{\prime}_{0} & \zeta_{0} 
\end{array}\right) y_{d} v_{d} \phi_{0} \; , \\
M_{e} & = & \left( \begin{array}{ccc}
0 & -(1-i) \phi_{0} \psi^{\prime}_{0} & \phi_{0} \psi^{\prime}_{0} \\
(1+i) \phi_{0} \psi^{\prime}_{0} & -3 \psi_{0} N_{0} & \phi_{0} \psi^{\prime}_{0} \\
0 & 0 & \zeta_{0} 
\end{array}\right) y_{d} v_{d} \phi_{0} \; . 
\end{eqnarray}

In the neutrino sector, the superpotential leads to the following Dirac neutrino mass matrix,
\begin{equation}
M_{D} = \left( \begin{array}{ccc}
2\xi_{0} + \eta_{0} & -\xi_{0} & -\xi_{0} \\
-\xi_{0} & 2\xi_{0} & -\xi_{0} + \eta_{0} \\
-\xi_{0} & -\xi_{0}+\eta_{0} & 2\xi_{0} 
\end{array}\right) \zeta_{0} \zeta^{\prime}_{0} v_{u} 
\equiv h_{D} v_{u} \; ,
\end{equation}
and the RH neutrino Majorana mass matrix,
\begin{equation}
M_{RR} = \left( \begin{array}{ccc}
1 & 0 & 0 \\
0 & 0 & 1 \\
0 & 1 & 0 
\end{array}\right) s_{0} \Lambda \; .
\end{equation}

We note that the complex CG coefficients appear in the product rules that involve the spinorial representations, ${\bf 2, \; 2^{\prime}, \; 2^{\prime\prime}}$. Because $(T_{1},T_{2})$ transform as the spinorial representation ${\bf 2}$, the charged fermion mass matrices, $M_{u}, \; M_{d}, \; M_{e}$, are complex. On the other hand, the neutrino involve only the vectorial-like representations, ${\bf 1,  \; 1^{\prime}, 1^{\prime\prime}, 3}$, and thus the neutrino Dirac and Majorana mass matrices are real and thus non-CP violating.

Note that the Dirac neutrino mass matrix, $M_{D}$, is diagonalizable by the TBM mixing matrix,
\begin{equation}
\label{eq:MDdiag}
V^{\dagger} K^{1/2} M_{D} K^{1/2} V = M_{D}^{\mbox{\tiny diag}} \quad \mbox{and} \quad V = U_{\mbox{\tiny TBM}} \; ,
\end{equation} 
where all elements in the diagonal matrix $M_{D}^{\mbox{\tiny diag}}$ are real, and $K$ is a diagonal phase matrix. 
The effective neutrino mass matrix, $M_{\nu}$, is obtained upon the seesaw mechanism taking place, 
\begin{equation}
M_{\nu} = -M_{D} M_{RR}^{-1} M_{D}^{T} \; ,
\end{equation}
which is diagonalizable by the tri-bimaximal mixing matrix,
\begin{equation}
U_{TBM}^{T} M_{\nu} U_{TBM} = \mbox{diag}( (3\xi_{0} + \eta_{0})^{2}, \eta_{0}^{2}, -(-3\xi_{0}+\eta_{0})^{2}) \frac{(\zeta_{0} \zeta_{0}^{\prime} v_{u})^{2} }{ s_{0}\Lambda} \; .
\end{equation}
One special property of $M_{\nu}$ is that it is form diagonalizable~\cite{Chen:2009um}. In other words, regardless of the values of $\xi_{0}$ and $\eta_{0}$, $M_{\nu}$ is always diagonalized by the tri-bimaximal mixing matrix, $U_{\mbox{\tiny TBM}}$. While the diagonalization matrix in the neutrino sector is independent of the model parameters in $M_{\nu}$, the mass eigenvalues for effective neutrinos are functions of the model parameters. Given that there are three absolute masses which are determined by two model parameters, there is one sum rule among the three absolute masses. Specifically, the sum rules are,
\begin{equation}
\biggl| |\sqrt{m_{1}}| + |\sqrt{m_{3}}| \biggr| = 2 | \sqrt{m_{2}}| \; ,  
\end{equation}
for $(3\xi_{0} + \eta_{0})(3\xi_{0} - \eta_{0}) >0$, and
\begin{equation}
\biggl| |\sqrt{m_{1}}| - |\sqrt{m_{3}}| \biggr| = 2 | \sqrt{m_{2}}| \; ,  
\end{equation}
for $(3\xi_{0} + \eta_{0})(3\xi_{0} - \eta_{0}) < 0$. 
If we express the $\Delta m_{\odot}^{2} \equiv m_{2}^{2} - m_{1}^{2}$ for the solar neutrinos and $\Delta m_{\mbox{\tiny atm}}^{2} \equiv m_{3}^{2} - m_{1}^{2}$ for the atmospheric neutrinos in terms of the model parameters, $\xi_{0}$ and $\eta_{0}$, they are given by,     
\begin{eqnarray}
m_{2}^2 - m_{1}^{2} & = & (\eta_{0}^{4} - (3\xi_{0} + \eta_{0})^{4}) \frac{(\zeta_{0} \zeta_{0}^{\prime} v_{u})^{2}}{s_{0}\Lambda} \; , \\
m_{3}^{2} - m_{1}^{2} & = & -24 \eta_{0} \xi_{0}( 9 \xi_{0}^{2} + \eta_{0}^{2})  \frac{(\zeta_{0} \zeta_{0}^{\prime} v_{u})^{2}}{s_{0}\Lambda} \; . 
\end{eqnarray}
Due to the presence of the matter effect in the solar neutrino oscillation, it is known that $\Delta m_{\odot}^{2} > 0$. This leads to the condition that $(\xi_{0} \cdot \eta_{0}) < 0$, which subsequently implied that $m_{3}^{2} - m_{1}^{2} > 0$. As a result, the normal hierarchy is predicted in our model.

We comment that the down-type quark and charged lepton mass matrices, $M_{d,e}$, are non-diagonal as a result of the Georgi-Jarlskog relations. This leads to a sizable $(12)$ mixing in $M_{e}$. In our model, the Cabibbo angle is predicted to be $\theta_{c} \simeq \sqrt{m_{d}/m_{s}}$. Similarly, we have the $(12)$ mixing angle in the charged lepton sector, $\theta_{12}^{e} \simeq \sqrt{m_{e} / m_{\mu}}$, which can be rewritten, using the Georgi-Jarlskog relations at the GUT scale, as $\theta_{12}^{e} \simeq \theta_{c} / 3$. This leads to a correction to the TBM mixing pattern in terms of the Cabibbo angle. Specifically, our model predicts a non-zero $\theta_{13}$, 
\begin{equation}
\theta_{13} \simeq \frac{\theta_{c}}{3\sqrt{2}} \; ,
\end{equation}
and the predicted solar mixing angle is given in  terms of the following quark-lepton complementarity sum rules~\cite{Antusch:2005kw},
\begin{equation}
\tan^{2} \theta_{\odot} \simeq \tan^{2} \theta_{\odot}^{\mbox{\tiny TBM}} + \frac{1}{2} \theta_{c} \cos\delta_{\ell}^{\mbox{\tiny CP}} 
= \frac{1}{2} + \frac{1}{2} \theta_{c} \cos\delta_{\ell}^{\mbox{\tiny CP}}  
\; .
\end{equation}

\section{Numerical Predictions}
\label{sec:Numeric}

The predictions for the charged fermion mass matrices in our model are schematically parametrized in terms of 7 parameters as~\cite{Chen:2007afa,Chen:2009gf},
\begin{eqnarray}
\frac{M_{u}}{y_{t} v_{u}} & = & \left( \begin{array}{ccccc}
i g & ~~ &  \frac{1-i}{2}  g & ~~ & 0\\
\frac{1-i}{2} g & & g + (1-\frac{i}{2}) h  & & k\\
0 & & k & & 1
\end{array}\right)  , \\
\frac{M_{d}, \; M_{e}^{T}}{y_{b} v_{d} \phi_{0}\zeta_{0}}  & = &  \left( \begin{array}{ccccc}
0 & ~~ & (1+i) b & ~~ & 0\\
-(1-i) b & & (1,-3) c & & 0\\
b & &b & & 1
\end{array}\right)  \; .
\end{eqnarray}
With $b \equiv \phi_{0} \psi^{\prime}_{0} /\zeta_{0} = 0.00304$, $c\equiv \psi_{0}N_{0}/\zeta_{0}=-0.0172$,  $k \equiv y^{\prime}\psi_{0}\zeta_{0}=-0.0266$, $h\equiv \phi_{0}^{2}=0.00426$ and $g \equiv \phi_{0}^{\prime 3}= 1.45\times 10^{-5}$, the following mass ratios are obtained, 
$m_{d}: m_{s} : m_{b} \simeq \theta_{c}^{\scriptscriptstyle 4.7} : \theta_{c}^{\scriptscriptstyle 2.7} : 1$, 
$m_{u} : m_{c} : m_{t} \simeq  \theta_{c}^{\scriptscriptstyle 7.5} : \theta_{c}^{\scriptscriptstyle 3.7} : 1$, 
with $\theta_{c} \simeq \sqrt{m_{d}/m_{s}} \simeq 0.225$. We have also taken $y_{t}/\sin\beta = 1.25$ and $y_{b}\phi_{0} \zeta_{0} / \cos\beta  \simeq 0.011$, which fit to $m_{t}$ and $m_{b}$, where $\tan\beta \equiv v_{u}/v_{d}$. 

In the numerical results quoted below, we have included the renormalization group corrections. As a consequence of the Georgi-Jarlskog (GJ) relations, realistic charged lepton masses are obtained. With these input parameters, the complex CKM matrix is,
\begin{eqnarray}
\left( \begin{array}{ccc}
0.974e^{-i 25.4^{o}} & 0.227 e^{i23.1^{o}} & 0.00412e^{i166^{o}} \\
0.227 e^{i123^{o}} & 0.973 e^{-i8.24^{o}} & 0.0412 e^{i180^{o}} \\
0.00718 e^{i99.7^{o}} & 0.0408 e^{-i7.28^{o}} & 0.999
\end{array}\right). 
\end{eqnarray}
Values for all $|V_{\mbox{\tiny CKM}}|$ elements are consistent with current experimental values~\cite{Amsler:2008zzb} except for $|V_{td}|$, the experimental determination of which has large hadronic uncertainty.  The predictions of our model for the angles in the unitarity triangle are, 
$\beta = 23.6^{o}$ ($\sin2\beta  =  0.734$), $\alpha = 110^{o}$, and $\gamma = \delta_{q} = 45.6^{o}$, 
(where $\delta_{q}$ is the CP phase in the standard parametrization), and they agree with the direct measurements within $1\sigma$ of BaBar and $2\sigma$ of Belle.  Potential direct measurements for these parameters at the LHCb and SuperB Factory can test our predictions. 

In the neutrino sector, with the following three input parameters (among which only two are independent),
\begin{eqnarray}
\xi_{0}\zeta_{0} \zeta_{0}^{\prime} = -0.00791 \; , \quad 
\eta_{0} \zeta_{0} \zeta_{0}^{\prime} = 0.01707\; , \quad
s_{0} \Lambda = 10^{12} \; \mbox{GeV} \; ,
\end{eqnarray}
the predictions for the mass square differences are
\begin{equation}
\Delta m_{32}^{2} = 2.54 \times 10^{-3} \; \mbox{eV}^{2} \; , 
\Delta m_{21}^{2} = 7.59 \times 10^{-5} \; \mbox{eV}^{2} \; ,
\end{equation}
with the three absolute masses being
\begin{equation}
m_{1} = 0.0156 \; \mbox{eV}, \; m_{2} = 0.179 \; \mbox{eV},  \; m_{3}  = 0.0514 \; \mbox{eV} \; ,
\end{equation}
and the two Majorana phases being 
\begin{equation}
\alpha_{21} = \pi, \; \alpha_{31} = 0 \; .
\end{equation}
The exact tri-bimaximal mixing pattern is corrected due to the presence of non-diagonal charged lepton mass matrix, which gives,
\begin{equation}
U_{e,L} = \left( \begin{array}{ccc}
0.838 e^{-i178^{o}} & 0.543 e^{-i173^{o}} & 0.0582 e^{i138^{o}}  \\
0.362 e^{-i3.99^{o}} & 0.610 e^{-i173^{o}} & 0.705 e^{i3.55^{o}}  \\
0.408 e^{i180^{o}} & 0.577 & 0.707
\end{array}\right) \; .
\end{equation}
The leptonic mixing parameters are predicted,
\begin{equation}
\sin^{2} 2\theta_{23} =1, \; \sin^{2} \theta_{12} = 0.296, \; \sin^{2}2\theta_{13} = 0.013, \; \delta_{\ell}^{\mbox{\tiny CP}} = 227^{o} \; .
\end{equation}
These results agree with the experimental values within $2\sigma$. We also note that our prediction of $\delta_{\ell}^{\mbox{\tiny CP}} = 227^{o}$ compares well with the SuperK best fit value of $\delta_{\ell}^{\mbox{\tiny CP}} = 220^{o}$~\cite{superK2010}.

\section{Leptogenesis}
\label{sec:Leptg}

The presence of the leptonic CP phase opens up the possibility of leptogenesis~\cite{Fukugita:1986hr,Chen:2007fv} which can gives rise to the cosmological matter antimatter asymmetry in the universe. Due to the additional parameters in high energy theory associated with the right-handed neutrino sector, nevertheless, it is generally not possible to connect leptogenesis to low energy parameters in neutrino oscillation. On the other hand, as the leptonic Dirac CP phase is the only non-vanishing CP violating predicted in our model, there is a strong correlation between leptogenesis and CP violation in neutrino oscillation in our model. (Another framework where a correlation can be established between leptogenesis and low energy CP violation is the minimal left-right symmetric model with spontaneous CP violation where there is only one physical phase in the lepton sector~\cite{Chen:2004ww}.)

It has been pointed out~\cite{Jenkins:2008rb} that in models in which the neutrino tri-bimaximal mixing pattern is generated by an underlying finite group family symmetry, such as $A_{4}$, there is no leptogenesis that can be generated, even in the presence of the flavor effects. We note that this results hold in the case of usual seesaw realization in which the RH neutrino masses are hierarchical. In the alternative seesaw realization in our model, the three RH neutrinos have near degenerate masses, leading to an enhanced, non-vanishing asymmetry through resonant leptogenesis~\cite{Pilaftsis:1997jf}, in the presence of the flavor effect. We  note that similar scenario has been considered in Ref.~\cite{Branco:2009by} in the context of $A_{4}$ symmetry. Nevertheless, Ref.~\cite{Branco:2009by} utilizes a different basis for the $A_{4}$ generators and the CP violating phases are not predicted. (Subsequent studies on leptogenesis in TBM models, see Ref.~\cite{Hagedorn:2009jy}.) Before discussing the asymmetry generation in our model, we first review the arguments for the vanishing asymmetry in the case of the usual seesaw realization.

In the Casas-Ibarra parametrization, the asymmetry due to the $i$-th RH neutrino, $N_{i}$, decay into a charged lepton of  $\alpha$ flavor is~\cite{Pascoli:2006ci},
\begin{equation}
\label{eq:FlavorAsym}
\epsilon_{i\alpha} = -\frac{3M_{i}}{16 \pi v^{2}} \frac{\mbox{Im} ( 
\sum_{\beta\rho} m_{\beta}^{1/2} m_{\rho}^{3/2} U_{\alpha\beta}^{\ast}
U_{\alpha\rho} R_{i\beta}R_{i\rho})}{
\sum_{\beta} m_{\beta} |R_{i\beta}|^{2}}   \; .
\end{equation}
Here the $R$ matrix is defined as 
\begin{eqnarray}
R =  vM^{-1/2} h U_{\mbox{\tiny MNS}} m^{-1/2} 
=  v M^{-1/2} (U_{\mbox{\tiny TBM}}^{T} h_{D}U_{e_{L}}) U_{\mbox{\tiny MNS}} m^{-1/2} \; ,
\end{eqnarray}
where $h$ is the Dirac Yukawa in the $M_{e}$ and $M_{\mbox{\tiny RR}}$ diagonal basis, 
$M = \mbox{diag}(M_{1},M_{2},M_{3})$ are the RH neutrino absolute masses, and 
$m = \mbox{diag}(m_{1},m_{2},m_{3})$ are the light neutrino absolute masses. 

Due to the hierarchy in the charged lepton masses, the Yukawa interactions involving three charged lepton flavors, $e, \; \mu, \; \tau$,  become equilibrium at temperatures around $10^{6}$, $10^{9}$, and $10^{12}$ GeV, respectively. If leptogenesis takes place at a scale above $10^{12}$ GeV, the Yukawa interactions involve the three lepton flavors are very weak and thus the three lepton flavors are indistinguishable. 
The total asymmetry is
\begin{equation}
\epsilon_{i} \equiv \sum_{\alpha} \epsilon_{i \alpha} = -\frac{3M_{i}}{16 \pi v^{2}} \frac{\mbox{Im} ( 
\sum_{\rho} m_{\rho}^{2} R_{i\rho}^{2})}{
\sum_{\beta} m_{\beta} |R_{i\beta}|^{2}} \; .
\end{equation}
However, if leptogenesis occurs at a scale below $10^{12}$ GeV, the one flavor approximation is no longer valid, and one needs to trace the asymmetries associated with different flavors individually. 
From Eq.~\ref{eq:FlavorAsym}, it is clear that in order to have a non-vanishing lepton number asymmetry, the following conditions must be satisfied: 
({\it i}) in the absence of flavor effects, the $R$ matrix must be complex and non-diagonal; 
({\it ii}) in the presence of flavor effects, the $R$ matrix must be non-diagonal. 

In the usual seesaw realization, with the following superpotential in the neutrino sector, 
\begin{equation}
\mathcal{W}_{\nu}^{usual} = H_{5} \overline{F} N + NN (\xi + \eta) \; ,
\end{equation}
the resulting RH Majorana mass matrix $(M_{RR})$ and Dirac neutrino Yukawa matrix $(h_{D})$ are given by,
\begin{eqnarray}
M_{RR} & = & \left( \begin{array}{ccc}
2\xi_{0} + \eta_{0} & -\xi_{0} & -\xi_{0} \\
-\xi_{0} & 2\xi_{0} & -\xi_{0} + \eta_{0} \\
-\xi_{0} & -\xi_{0} + \eta_{0} & 2 \xi_{0}
\end{array}\right) \Lambda \; , \\
M_{D} & = & \left( \begin{array}{ccc}
1 & 0 & 0 \\
0 & 0 & 1 \\
0 & 1 & 0
\end{array}\right) v \equiv h_{D} v_{u} \; .
\end{eqnarray}
The RH Majorana mass matrix is diagonalized by the TBM mixing matrix,
\begin{equation}
U_{TBM}^{T} M_{RR} U_{TBM} = \mbox{diag} (3\xi_{0} + \eta_{0}, \eta_{0}, 3\xi_{0} - \eta_{0}) \Lambda \quad .
\end{equation} 
In the basis where $M_{RR}$ and $M_{e}$ are real and diagonal, the Dirac neutrino Yukawa matrix reads,
\begin{equation}
h = U_{TBM}^{T} h_{D} U_{e_{L}} \; .
\end{equation}
Thus the $R$ matrix is given by
\begin{eqnarray}
R & = &  v M^{-1/2} (U_{\mbox{\tiny TBM}}^{T} h_{D}U_{e_{L}}) U_{\mbox{\tiny MNS}} m^{-1/2} \\
& = &  v M^{-1/2} U_{\mbox{\tiny TBM}}^{T} h_{D} U_{e_{L}} U_{e_{L}}^{\dagger} U_{\mbox{\tiny TBM}} m^{-1/2} 
= v M^{-1/2} U_{\mbox{\tiny TBM}}^{T} h_{D} U_{\mbox{\tiny TBM}} m^{-1/2} 
 =  v M^{-1/2} m^{-1/2} \; . \nonumber
\end{eqnarray}
Clearly $R$ is a real and diagonal matrix, leading to a vanishing lepton number asymmetry, even in the presence of the flavor effects.

In our seesaw realization, the RH Majorana neutrino mass matrix is given by,
\begin{equation}
M = I_{3\times 3} \cdot s_{0} \Lambda \; , \quad U_{\nu,R} = \left(\begin{array}{ccc}
1& 0 & 0 \\
0 & 1/\sqrt{2} & -i/\sqrt{2} \\
0 & 1/\sqrt{2} & i/\sqrt{2} \\
\end{array}\right) \; ,
\end{equation}
thus the high energy orthogonal matrix, $R$, takes the following form,
\begin{equation}
R = v M^{-1/2} U_{\nu,R} M_{D} U_{\mbox{\tiny TBM}} m^{-1/2} \; , 
\end{equation}
leading to a real, non-diagonal $R$ matrix with non-zero off-diagonal elements in the $(12)$ block. 

In the limit of three exact degenerate RH neutrino masses, the asymmetry vanishes. Nonetheless, the renormalization group equations give rise to small mass splitting among the RH neutrino masses, leading to an enhancement of the asymmetry through the self-energy diagram. The asymmetry associated with the $\alpha$ flavor due to the self-energy diagram is~\cite{Pascoli:2006ci},
\begin{equation}
\epsilon_{i}^{\alpha} = -\sum_{j  \ne i}  \frac{\Gamma_{j}}{M_{j}} S_{ij} I_{ij}^{\alpha} \; ,
\end{equation}
where the total decay width, $\Gamma_{j}$ of the $j$-th right-handed neutrino, $N_{j}$, is given by,
\begin{equation}
\Gamma_{j} = \frac{1}{8\pi} (hh^{\dagger})_{jj} M_{j} \; ,
\end{equation} 
the parameter $S_{ij}$ characterizes the resonance enhancement factor,
\begin{equation}
S_{ij} = \frac{M_{i}M_{j} \Delta M_{ij}^{2}}{(\Delta M_{ij}^{2})^{2} + M_{i}^{2}\Gamma_{j}^{2}} \; ,
\end{equation}
where
\begin{equation}
\Delta M_{ij}^{2} = M_{j}^{2} - M_{i}^{2} \; .
\end{equation}
The asymmetry stored in the $\alpha$ flavor is proportional to the parameter, $I_{ij}^{\alpha}$, where
\begin{equation}
I_{ij}^{\alpha} = \frac{1}{(hh^{\dagger})_{ii} (hh^{\dagger})_{jj}} \frac{M_{i}M_{j}}{v_{u}^{4}} \sum_{\ell} (R_{i\ell} R_{j\ell} m_{\ell}) 
\sum_{t,s} \sqrt{m_{t}m_{s}} R_{it} R_{js} \mbox{Im} (U_{\alpha s} U_{\alpha t}^{\ast}) \; .
\end{equation}  
In terms of the mass splitting parameter $\delta_{ij}^{R}$ defined as, 
\begin{equation}
\delta_{ij}^{R} \equiv \frac{M_{j}}{M_{i}} - 1 \; ,
\end{equation}
the mass splitting due to the RG corrections is given by~\cite{Branco:2005ye}, 
\begin{eqnarray}
\Delta M_{ij}^{2} & = & 2 M_{i}^{2} \delta_{ij}^{R} \; ,
\\
\delta_{ij}^{R} & = & 2 (\hat{H}_{ii} - \hat{H}_{jj}) t \; , \quad \; t \equiv \frac{1}{16\pi^{2}} \ln \biggl( \frac{M_{\mbox{\tiny GUT}}}{M} \biggr) \; ,
\end{eqnarray}
where $\hat{H} = V^{T} (h_{D} h_{D}^{\dagger}) V$,  and $V = U_{\mbox{\tiny TBM}}$ as defined in Eq.~\ref{eq:MDdiag}.   
The GUT scale, $M_{\mbox{\tiny GUT}}$ is taken to be $\sim 2 \times 10^{16}$ GeV.

With the input parameters in the Yukawa sector, the $R$ matrix is predicted to be
\begin{equation}
R = \left( \begin{array}{ccc}
-0.816 & 0.577 & 0 \\
0.577 & 0.816 & 0 \\
0 & 0 & i 
\end{array} \right) \; .
\end{equation}
The total decay widths of the right-handed neutrinos, $N_{j} \; (j = 1,2,3)$, are given by,
\begin{eqnarray}
\Gamma_{1}  = 2.52 \times 10^{6} \; \mbox{GeV} \; , \; 
\Gamma_{2} = 4.15 \times 10^{6} \; \mbox{GeV} \; , \; 
\Gamma_{3} = 3.31 \times 10^{7} \; \mbox{GeV} \; .
\end{eqnarray}
The RG corrections lead to the following values for the mass splitting parameters,
\begin{equation}
\delta_{21}^{R} \simeq 3.1 \times 10^{-5} \; , \; 
\delta_{32}^{R} \simeq 1.72 \times 10^{-4} \; , \; 
\delta_{31}^{R} \simeq 2.03 \times 10^{-4} \; ,
\end{equation} 
giving rise mass splittings that are on the same order as the decay widths, as required in order to have the resonance enhancement. 
The resulting resonance enhancement factors are,
\begin{equation}
S_{12} = -0.276 \; , \quad
S_{13} = -0.0458 \; , \quad
S_{23} = -0.0539 \; .
\end{equation}
At $T\sim 10^{12}$ GeV, only the $\tau$ Yukawa interaction is in equilibrium, and we have 
\begin{equation}
-\epsilon_{i}^{\tau} = \epsilon_{i}^{e} + \epsilon_{i}^{\mu}, \; \mbox{for} \; (i = 1,2,3) \; .
\end{equation}  
It is thus suffice to consider $\epsilon_{i}^{\tau}$ only. 
As the non-vanishing off diagonal elements in the $R$ matrix appear in the $(12)$ block, the generation of the lepton number asymmetry is due to the decays of the right-handed neutrinos, $N_{1}$ and $N_{2}$. The contribution from $N_{3}$ decay is negligible. The Hubble expansion rate at the leptogenesis temperature, $T \sim M$ is 
\begin{equation}
H (T \simeq M) \sim \frac{6.8 \times \sqrt{g_{\ast}} * M^{2}}{M_{\mbox{\tiny Pl}}} \sim 10^{8} \; \mbox{GeV} \; ,
\end{equation}
where the relativistic degrees of freedom $g_{\ast} \simeq 229$, 
the right-handed neutrino decays are thus out-of-equilibrium as $\Gamma_{i} < H(T \simeq M)$.   
The resulting lepton number asymmetry stored in the $\tau$ flavor due to $N_{1}$ decay of,
\begin{equation}
\epsilon^{\tau}_{1} \sim -9.04 \times 10^{-7} \; ,
\end{equation}
and the asymmetry due to $N_{2}$ decay is 
\begin{equation}
\epsilon^{\tau}_{2} \sim -1.33 \times 10^{-7} \; ,
\end{equation}
which are of the right order of magnitude for sufficient amount of baryon number asymmetry, $\eta_{B} = n_{B} / n_{\gamma} \simeq 6.1 \times 10^{-10}$~\cite{Bennett:2003bz}.

\section{Conclusion}
\label{sec:conclude} 
In this note, we investigate leptogenesis in a SUSY $SU(5) \times T^{\prime}$ model, in which CP violation is entirely geometrical in origin due to the  complex Clebsch-Gordon coefficients in  $T^{\prime}$. Our model gives rise to realistic quark masses and CKM matrix elements, a near tri-bimaximal neutrino mixing pattern, as well as CP violating measures for all quarks  and leptons. In terms of 9 independent parameters in the Yukawa sector, the predicted values of the 22 observables agree with current experimental values, including the recent T2K and MINOS results. The correction to the tri-bimaximal mixing pattern is related to the Cabibbo angle, leading to interesting quark-lepton complementarity sum rules. In addition, the normal hierarchy is predicted for the neutrino mass ordering. Our predicted value for the leptonic Dirac CP phase is $\delta_{\ell}^{\mbox{\tiny CP}} = 227^{o}$, which is very close to the current best fit value from SuperK, $\delta_{\ell} = 220^{o}$. The non-zero $\delta_{\ell}^{\mbox{\tiny CP}}$ phase gives rise to sufficient amount of lepton number asymmetry, in the presence of the flavor effect. As this is the only non-vanishing CP violating phase in the lepton sector, there is a direct connection between leptogenesis and CP violation in neutrino oscillation in our model. 

\section*{Acknowledgements}
KTM thanks Kaladi Babu for useful discussions.  
The work of M-CC was supported, in part, by the National Science Foundation under grant No. PHY-0970173. The work of KTM was supported, in part, by the Department of Energy under Grant No. DE-FG02-04ER41290.

\end{document}